\documentclass[11pt,a4paper,aps,notitlepage]{revtex4-1} 

\usepackage{graphicx}
\usepackage{amsmath}
\usepackage{amsfonts}
\usepackage{amssymb}
\usepackage{subfig}
\usepackage[justification=justified]{caption}
\usepackage{verbatim}

\usepackage[colorlinks=true]{hyperref}
\usepackage[all]{hypcap} 
\usepackage{cleveref}

\begin{document} 
\title{Interlayer excitons and Band Alignment in MoS$_2$/hBN/WSe$_2$ van der Waals Heterostructures}

\author{Simone Latini}
\email{simola@fysik.dtu.dk}
\affiliation{Center for Atomic-scale Materials Design, Department of Physics, Technical University of Denmark, DK - 2800 Kgs. Lyngby, Denmark}

\author{Kirsten T. Winther}
\affiliation{Center for Atomic-scale Materials Design, Department of
Physics, Technical University of Denmark, DK - 2800 Kgs. Lyngby, Denmark}

\author{Thomas Olsen}
\affiliation{Center for Atomic-scale Materials Design, Department of
Physics, Technical University of Denmark, DK - 2800 Kgs. Lyngby, Denmark}

\author{Kristian S. Thygesen}

\affiliation{Center for Atomic-scale Materials Design, Department of Physics, Technical University of Denmark, DK - 2800 Kgs. Lyngby, Denmark}
\affiliation{Center for Nanostructured Graphene, Technical University of Denmark, DK - 2800 Kgs. Lyngby, Denmark}

\begin{abstract}
Van der Waals heterostructures (vdWH) provide an ideal playground for exploring light-matter interactions at the atomic scale. In particular, structures with a type-II band alignment can yield detailed insight into free carrier-to-photon conversion processes, which are central to e.g. solar cells and light emitting diodes. An important first step in describing such processes is to obtain the energies of the interlayer exciton states existing at the interface. Here we present a general first-principles method to compute the electronic quasi-particle (QP) band structure and excitonic binding energies of incommensurate vdWHs. The method combines our quantum electrostatic heterostructure (QEH) model for obtaining the dielectric function with the many-body GW approximation and a generalized 2D Mott-Wannier exciton model. We calculate the level alignment together with intra and interlayer exciton binding energies of bilayer MoS$_2$/WSe$_2$ with and without intercalated hBN layers, finding excellent agreement with experimental photoluminescence spectra. Comparison to density functional theory calculations demonstrate the crucial role of self-energy and electron-hole interaction effects.    
\end{abstract}
\maketitle

The use of two-dimensional (2D) transition metal dichalcogenides\cite{Wang2012,Fai2010,Splendiani2010,Ramasubramaniam2012} as fundamental building blocks in (opto)electronics has proved highly promising for the construction of ultrathin high performance devices\cite{Jariwala2014,Lopez2013,Ross2014,Pospischil2014,Steinhoff2015,Massicotte2016}. By reassembling different 2D crystals into van der Waals heterostructures, designer materials with new and tailored properties can be made\cite{Geim2013,Terrones2013,Britnell2013,Withers2015,Bernardi2013,Massicotte2016,Ma2016,Hong2014}.
 As for 2D monolayers\cite{Qiu2013,Ugeda2014,Keliang2014,Zhang2014,Hybertsen2013,Molina2016}, the optical properties of few-layer van der Waals heterostructures are strongly influenced by excitonic effects\cite{Bernardi2013,Massicotte2016,Ma2016} as a consequence of the weak screening of the electron-hole interaction\cite{Latini2015}.  In addition to the intralayer excitons localized in the constituent monolayers, vdWHs can host more complex types of excitons with electrons and holes residing in distinct layers, so-called (spatially) indirect excitons or interlayer excitons. 
Because of the spatial charge separation, interlayer excitons posses longer electron-hole recombination lifetimes\cite{Rivera2015,Palummo2015} than intralayer excitons, which make them ideal candidates for realization of bosonic many-particle states like Bose-Einstein condensates\cite{Fogler2014}.  Moreover, interlayer excitons are believed to play a central role in the charge separation process following photoabsorption in solar cells or photodetectors\cite{Lee2014,Skinner2016}. Of key importance to this process is the exciton binding energy which quantifies the strength with which the electron and hole are bound together. Due to the larger electron-hole separation, interlayer excitons are expected to have lower binding energies than intralayer excitons. However, a detailed understanding of interlayer excitons in vdWHs is still lacking mainly  because of the highly non-local nature of the dielectric function of 2D materials which makes  screening less effective at larger distances. This is in fact the origin of the non-hydrogenic Rydberg series in 2D semiconductors and the non-degeneracy of 2D excitons with different angular momentum quantum numbers\cite{Chernikov2014, Keliang2014, Ye2014, Olsen2016}. The understanding of excitonic effects alone, however, is not sufficient for device engineering, where the knowledge of the alignment of the electronic bands of the vdWH is also crucial. Several experimental investigations have shown, for example, that an underlying type-II band alignment is required for the formation of interlayer excitons\cite{Fang2014,Rivera2016,Heo2015}. However, experimental data has not been supported by consistent theoretical studies yet.  It is well known that density functional theory (DFT) calculations are problematic when it comes to prediction of band gaps and band alignment at interfaces and do not take excitonic effects into account. An important deficit of the DFT approach, in addition to its general tendency to underestimate band gaps, is its failure to describe image charge renormalization effects that shift the energy levels of a 2D semiconductor\cite{Huser2013} or molecule\cite{Garcia2009,Neaton2006} when adsorbed on a polarizable substrates. Ideally, one should employ many-body perturbation theory like the GW approximation to obtain reliable band energies and the Bethe Salpeter Equation (BSE) for exciton binding energies. However, the computational cost of such methods make them unfeasible for vdWHs containing more than a few lattice matched 2D crystals.  

Here, we show how to overcome these limitations, by means of our recently developed quantum electrostatic heterostructure (QEH) approach\cite{Andersen2015} and accurately calculate interlayer exciton binding energies and electronic bands of vdWHs.  For the excitons, the QEH allows us to calculate the the screened electron-hole interaction to be used in a generalized 2D Mott-Wannier model\cite{Latini2015}. For the band energies we use the QEH to modify isolated layer G$_0$W$_0$ calculations by including the effect of interlayer screening on the electronic levels. Remarkably we are able to predict band positioning in a vdWH at the cost of, at most, $N$-G$_0$W$_0$ monolayer calculations, with $N$ the number of layers in the stack. 

In this letter we apply our method to the case of MoS$_2$-WSe$_2$ bilayers intercalated with a varying number of h-BN layers. The MoS2/hBN/WSe2 represents a prototypical type-II heterostructure. Its well defined atomic structure  and the possibility of varying the thickness of the hBN spacer and the relative orientation of the photoactive layers, makes it an ideal platform for studying light-matter processes on atomic length scales. At the same time, the incommensurate nature of the van der Waals interfaces  presents a great  challenge for ab-initio calculations. Nevetheless, we show that our QEH-based methodology allows us to efficiently simulate the electronic structure of MoS$_2$/hBN/WSe$_2$, including excitonic and self-energy effects, and accurately reproduce experimental photoluminescence spectra.\\

\begin{figure*}[!b]
  \includegraphics[width = \linewidth]{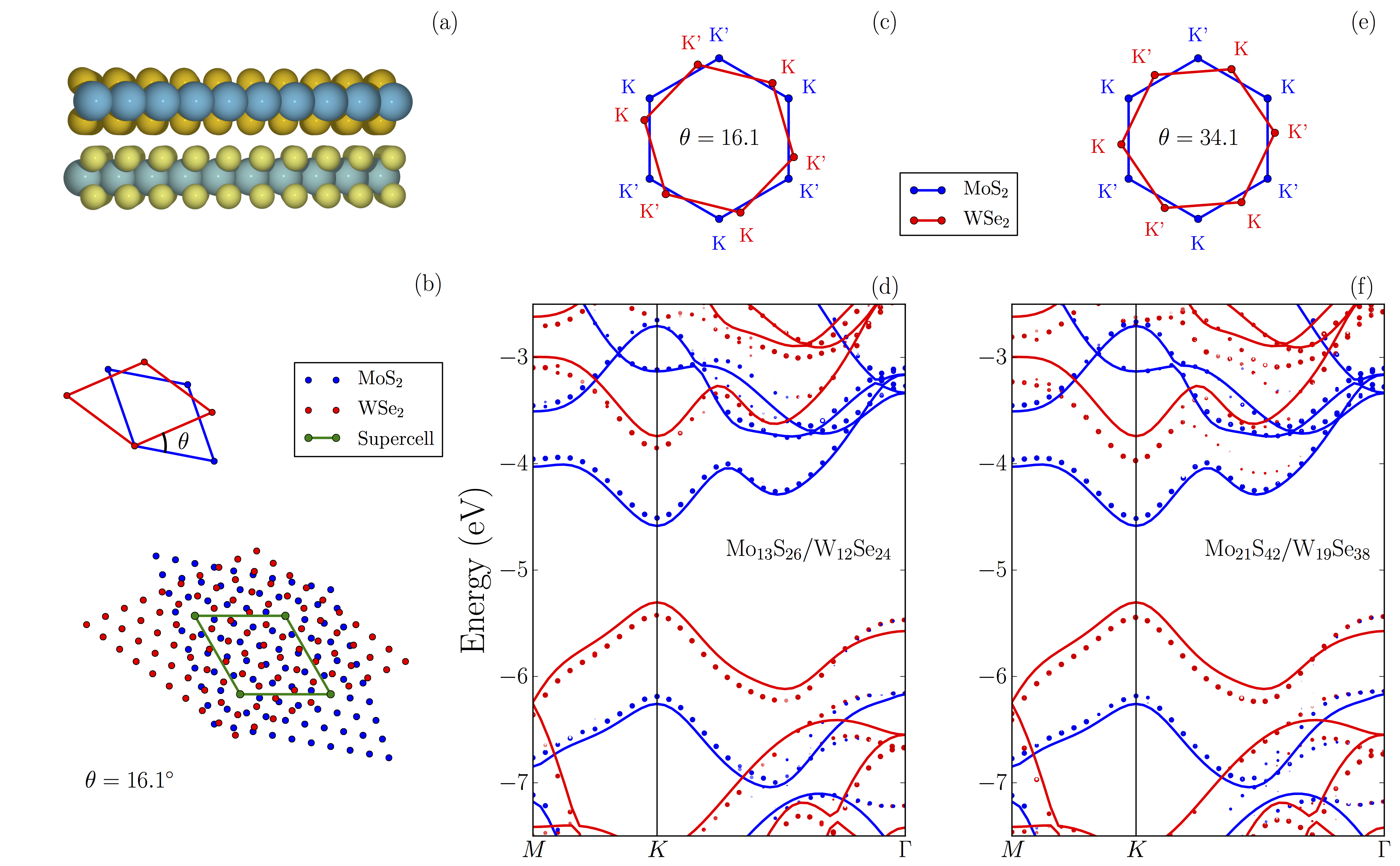}
   \caption{Panel (a) cartoon of the MoS$_2$-WSe$_2$ bilayer system. Panel (b) Representation of real space primitive and super-cell for $\theta=16.1^{\circ}$. Panels (c) and (e) illustrate how the BZs are twisted and differ in size for the two different alignment angles $\theta$, $16.1^{\circ}$ (a) and $34.4^{\circ}$ respectively. The unfolded LDA band structure with respect to vacuum for MoS$_2$-WSe$_2$ bilayers with for $\theta=16.1^{\circ}$ and $\theta=34.4^{\circ}$ are plotted in panel (d) and (f) respectively. Circles are used for the unfolded bilayer bands, while continuous lines are used for the isolated layers. The bands are colored in blue or red based on the character of the band, i.e. if it either belong to the MoS$_2$ or WSe$_2$ layer. More information about the color scheme are given in methods section. For comparison the isolated layers LDA bands are shown with continuous lines. For simplicity no spin-orbit coupling is included in the electronic bands.}
 \label{fig:bands_rotation}
\end{figure*}

The main requirement for the existence of interlayer exciton is that the bottom of the conduction band and the top of the valence band in a van der Waals stack are located in two different layers. As shown in the following, this is the case of MoS$_2$-WSe$_2$ based heterostructures. Because of lattice mismatch (see \cref{tab:geom}), MoS$_2$ and WSe$_2$ form incommensurable heterostructures and therefore realistic band structure calculations require the use of relatively large in-plane supercells as illustrated in panel (b) of \cref{fig:bands_rotation}. By rotating the layers with respect to each other, not only can the dimension of the supercell be reduced, but we can also mimic more closely the experimental situation where the alignment angle between the layers is not controlled. Following the procedure described by Komsa et al\cite{Komsa2013}, we set up the MoS$_2$-WSe$_2$ bilayer for two different alignment angles, specifically  $\sim 16.1^\circ$ and $\sim 34.4^\circ$, so that each layer is strained by less than $1\%$. To be able to compare the band structure of the bilayer with the ones of the constituent isolated monolayers, we unfold the electronic bands of the supercell to the ones of the primitive  MoS$_2$ and WSe$_2$ cells. This is done by following the method described in Ref.~\citenum{Popescu2012}. We stress that, because of the lattice mismatch and a non-zero alignment angle, the first Brillouin zones (BZ) of the two materials are different in size and rotated with respect to each other as shown in panel (c) and (e) of \cref{fig:bands_rotation}. This implies that the unfolding of the bands has to be performed accordingly.

\begin{figure*}[!b]
\includegraphics[width = .9\linewidth]{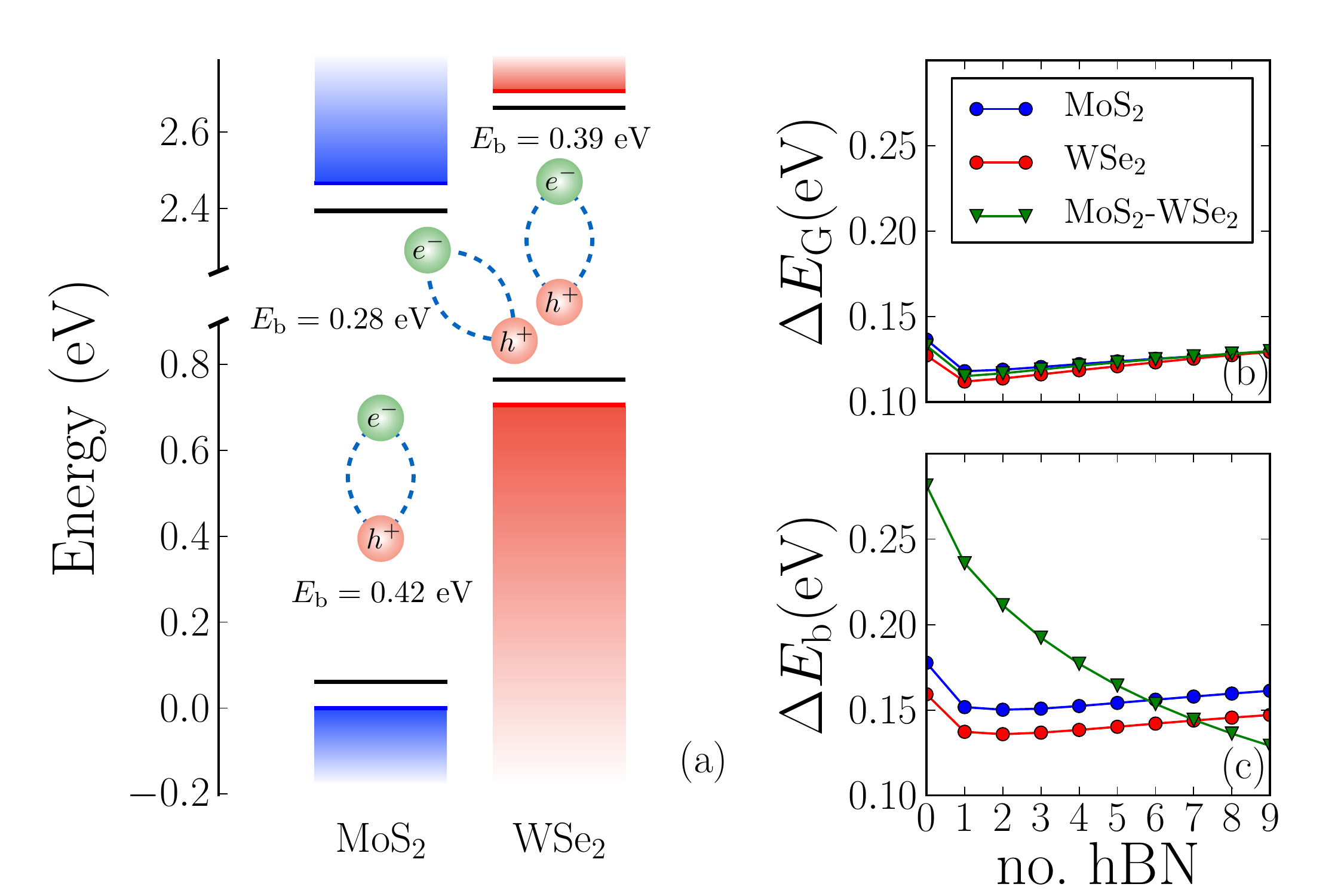}
  	  \caption{(a) Band alignment diagram for bilayer MoS$_2$/WSe$_2$ calculated with G$_0$W$_0$ in the isolated layers condition (colored bars) and with the G$_0\Delta$W$_0$ approximation (black lines) to account for the effect of interlayer screening.  In the same panel we depicted the intra and interlayer excitons in MoS$_2$/WSe$_2$ with their respective binding energies. Variation in intra and interlayer gaps (b) and exciton binding energies (c) as a function of intercalated BN layers.}
  \label{fig:model_sketch}
\end{figure*}

The unfolded band structures, aligned with respect to vacuum level, for the two different bilayers are shown as circles in \cref{fig:bands_rotation} (d) and (f) and compared to the isolated monolayers bands in continuous lines. The band structures have been calculated using the local density approximation (LDA) as described in the Methods section and for simplicity the effect of spin-orbit coupling is not included here.
%
%
%
%
%
%
We can clearly confirm that MoS$_2$ and WSe$_2$ form a type II heterostructure, where the top of the valence band is localized on the WSe$_2$ layer, while the bottom of the conduction band belongs to MoS$_2$. This implies that MoS$_2$/WSe$_2$ can host interlayer excitons as sketched in \cref{fig:model_sketch} (a). Furthermore, no significant difference in the band structure emerges for the two different alignment angles and thus we can conclude that the band structure of the bilayer is independent of the alignment angle. When comparing to the bands of the isolated monolayer, we can distinguish two main effects, as thoroughly shown in the Supporting Information: the effect of interlayer hybridization, only around the $\Gamma$ point, and a layer-dependent shift in energy throughout the Brillouin zone. For the latter, we observe the MoS$_2$ bands to be shifted up in energy while the WSe$_2$ bands are shifted down, with a consequent increase in indirect gap of $0.21$ eV relative to the vacuum level-aligned isolated layers. Such an asymmetric shift is a clear signature of the formation of a dipole at the interface of the two layers as a consequence of charge rearrangement induced by the misalignment between the band gap center of the two materials. To summarize, since hybrization is minimal and the charge transfer effect, if needed, can be accounted for just by adding a constant shift, we learn that the bilayer bands around the $K$-point can be directly obtained as a superposition of the constituent isolated monolayers bands. This result is consistent with the findings in refs.~\citenum{Zande2014,Liu2014} which show that even for the MoS$_2$ homobilayer, expected to be particularly dependent on hybridization as it is possible to choose twisting angles corresponding to the AA and AB stacking, the band edges at the $K$-valley are largely unaltered. We stress that the band edges at the $K$-point are the relevant ones for the formation and radiative recombination (photoluminescence) of interlayer excitons.

\begin{figure*}[!t]
  \includegraphics[width = \linewidth]{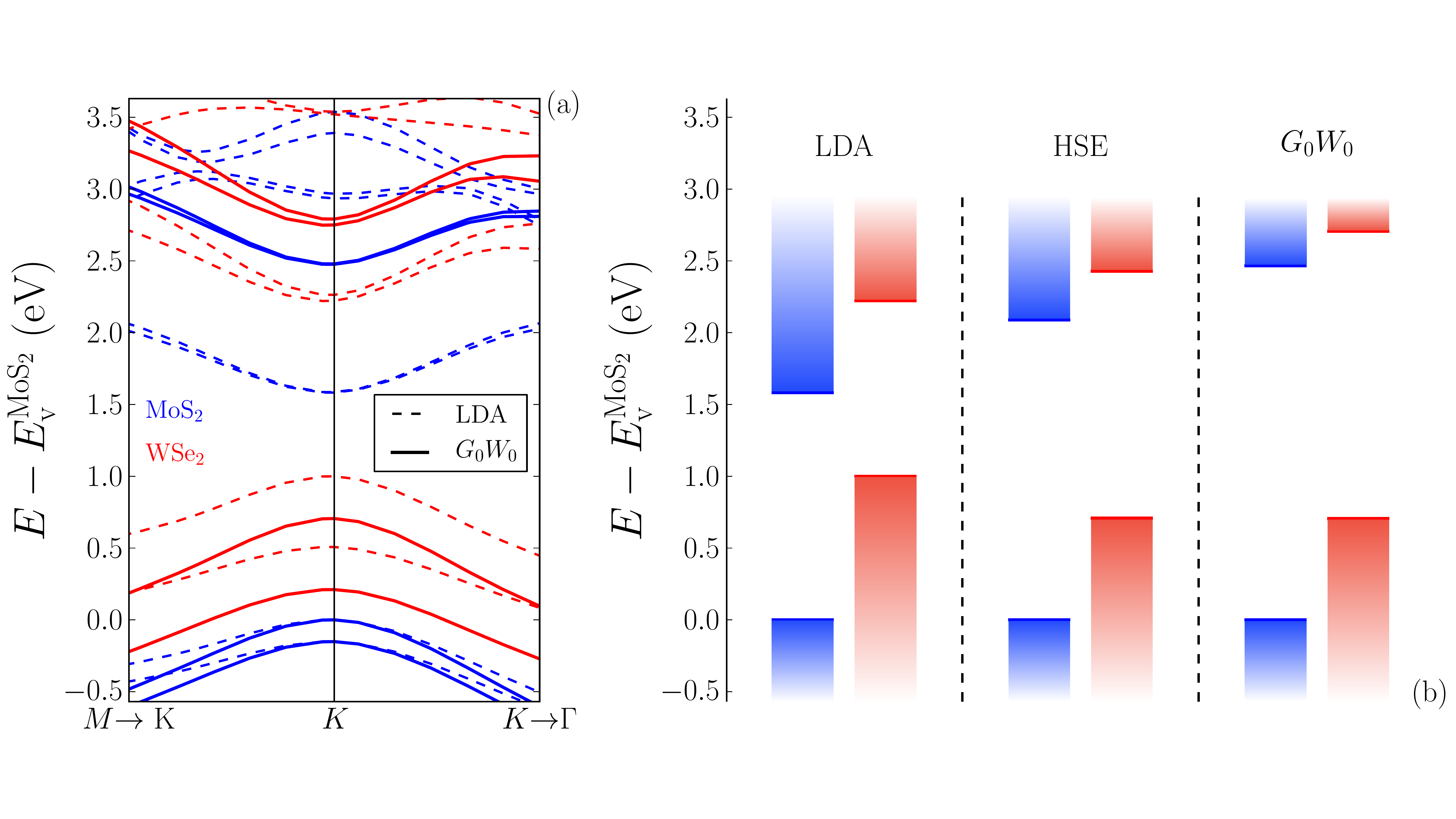}
   \caption{Panel (a) LDA and G$_0$W$_0$ band structures for the isolated layers. The panel illustrates a restricted part of the BZ around the K-point. The color code is the same as for \cref{fig:bands_rotation}, i.e. blue for MoS$_2$ and red for WSe$_2$. Panel (b) band edges within different approximations. Because of the uncertainty of the G$_0$W$_0$ vacuum levels, all the bands are aligned with respect to the top of the valence band in MoS$_2$. For the HSE calculations the HSE06 hybrid functional is used. Spin-Orbit Effects are included.}
 \label{fig:bands_DFTvsGW}
\end{figure*}

While in terms of computational cost it would be advantageous to utilize LDA band structures for quantitative description of bands in vdWHs, their use is largely questionable when accuracy on band alignment and on band gaps is required. To illustrate the LDA failure, we compare the LDA and G$_0$W$_0$ electronic bands for the isolated MoS$_2$ and WSe$_2$ monolayers in panel (a) of \cref{fig:bands_DFTvsGW}. 
In both approaches we include the effect of spin-orbit coupling at a non-self-consistent level. While the qualitative picture of a type II band alignment is preserved within the G$_0$W$_0$ approximation,  the band alignment and band gaps predicted by LDA are wrong. This is even more evident in panel (b) of \cref{fig:bands_DFTvsGW}, where we directly show the band edges for the isolated MoS$_2$ and WSe$_2$ layers. One could possibly argue that the inaccuracy is due to the LDA exchange correlation functional rather than the DFT approach itself. For this reason we also calculated band edges using the HSE06 hybrid functional, which is known to perform well for band structures. However, as shown in panel (b), HSE is better than LDA but still not as accurate as G$_0$W$_0$. We thus conclude that the G$_0$W$_0$ approximation is essential to obtain a quantitatively correct description of the band gaps and band alignment. Importantly, we notice that while DFT predicts a rather large difference in band gap centers of around $1$ eV, G$_0$W$_0$ gives a much smaller difference of around $0.3$ eV. This, together with the increased band gaps predicted by G$_0$W$_0$, strongly indicates that the charge transfer and associated dipole shift of the bands, could be significantly overestimated by DFT-LDA. We thus conclude that the band structure of the heterobilayer around the K-point can be obtained by combining the G$_0$W$_0$ band structures of the isolated layers aligned with respect to a common vacuum level and corrected for image charge screening effects (see later).\\

The state of the art for describing excitonic effects from first principles is the many-body Bethe-Salpeter Equation (BSE)\cite{Salpeter1951,Strinati1984,Onida2002}. The solution of the BSE is, however, computationally demanding already at the monolayer level and practically impossible for incommensurate van der Waals heterostructures. However, it is well-known that, under well defined assumptions the excitonic many-body problem can be rephrased in terms of an effective hydrogenic Hamiltonian, the Mott-Wannier Hamiltonian\cite{GrossoPastori2000}, which gives a satisfactory description of several excitonic properties\citep{Cudazzo2010,Cudazzo2011,Pulci2012,Latini2015}. In the case of excitons in vdWHs, the motion of the electron and the hole is restricted to the in-plane direction. This is a direct consequence of the anisotropy of layered structures, which entails that the effective masses at the K-point in out-of-plane direction are much higher than the in-plane direction. With this consideration the hydrogenic Hamiltonian reduces to a 2D problem:
\begin{equation}
\label{eq:MW}
\left[-\frac{\nabla_{2D}^2}{2\mu_\textrm{ex}}+W({\bf r}_\parallel)\right]F({\bf r}_\parallel)=E_\textrm{b}F({\bf r}_\parallel),
\end{equation}
where $\mu_\textrm{ex}$ is the exciton effective mass  and $W({\bf r}_\parallel)$ is the electron-hole interaction energy. The exciton effective mass is evaluated as  $\mu_\textrm{ex}^{-1}=m_\textrm{e}^{-1}+m_\textrm{h}^{-1}$, where the hole and electron masses are calculated ab-initio and reported in \cref{tab:geom}.
\begin{table}[b]
\caption{Lattice parameters and effective masses. The latter are calculated at the point K of the BZ.}
\resizebox{0.5\textwidth}{!}{
\begin{tabular}{cccccc}
\hline
Material & $a_\textrm{MM}$ (\AA) & $a_\textrm{XX}$ (\AA) & $m_\textrm{h}$ & $m_\textrm{e}$ & $\mu_\textrm{intra}$\\ \hline
MoS$_2$ & 3.18 & 3.13 & 0.56 & 0.55 & 0.27 \\ 
WSe$_2$ & 3.30 & 3.31 & 0.48 & 0.44 & 0.23 \\ 
 \hline
\end{tabular}}
\label{tab:geom}
\end{table}

In the case of interlayer excitons \cref{eq:MW} is still valid. Indeed even if the electron and the hole are separated in the out-of-plane direction, their motion is still confined in their respective layers. On the other hand, this spatial separation affects the screened electron-hole interaction energy, as shown below. In the specific case of MoS$_2$/WSe$_2$ heterostructures, the electron  and hole effective masses have to be estimated from the WSe$_2$ and MoS$_2$ valence and conduction bands respectively. We found an interlayer exciton effective mass of $0.244$ a.u.

The Coulomb interaction between the electron and the hole in an exciton is highly sensitive to the dielectric properties of the material\cite{Komsa2012}. As we showed in our recent work \cite{Latini2015}, the dielectric screening of finite thickness vdWHs is strongly non local. Using our recently developed quantum-electrostatic heterostructure (QEH) model, we can determine the dielectric properties of these complex structures from first principles\cite{Andersen2015}.  Briefly, we first condense the dielectric response of each single layer into a ``dielectric building block'' consisting of the monopole and dipole components of the density response function of the isolated layer\cite{Yan2012}. Second, the dielectric building blocks are coupled together electrostatically by solving a Dyson-like equation in the discrete monopole/dipole basis in order to obtain the density response function for the whole structure. The underlying assumption of the QEH is that hybridization is weak enough that is does not affect the dielectric properties of the heterostructure. We have found that this approximation is surprisingly good.  

A collection of more than 50 dielectric building blocks together with the software for the electrostatic coupling can be found in Ref.~\citenum{database2015}

From the response function, the dielectric function of the heterostructure is determined and it can be used to obtain the screened electron-hole interaction energy:
\begin{equation}
\label{eq:W_q_vdWH}
W(q_\parallel) = \underline{\rho}_{\hspace{0.05cm}\textrm{e}}^{\intercal}(q_\parallel)~\underline{\underline{\epsilon}}^{-1}(q_\parallel)~\underline{\phi}_{\hspace{0.05cm}\textrm{h}}(q_\parallel),
\end{equation}
where $\underline{\rho}_{\hspace{0.05cm}\textrm{e}}$ ($\underline{\phi}_{\hspace{0.05cm}\textrm{h}}$) is the electron density (hole induced potential) vector expressed in a basis set of monopole/dipole densities (potentials). The basis set of induced densities and potentials is also used as (left and right) basis functions for representing $\underline{\underline{\epsilon}}^{-1}$ (see Ref.~\citenum{Andersen2015}). The underlying structure of a vector in the multipole basis can be readily understood. An arbitrary density vector $\underline{\rho}$ is represented as $\underline{\rho}^{\intercal}=[\rho_\textrm{1M},\rho_\textrm{1D},\rho_\textrm{2M},\rho_\textrm{2D},\cdots,\rho_\textrm{nM},\rho_\textrm{nD}]$ where $\rho_{i\alpha}$, with $\alpha=\textrm{M,D}$, is the induced monopole/dipole density at the layer $i$. A completely equivalent representation can be used for the potential. Now, for the specific case of an interlayer exciton in the MoS$_2$/WSe$_2$ bilayer, the electron density vector takes the form $\underline{\rho_\textrm{e}}^{\intercal}=[\rho_\textrm{1M},\rho_\textrm{1D},0,0]$, while the potential induced by the hole is $\underline{\phi_\textrm{h}}^{\intercal}=[0,0, \phi_\textrm{2M},\phi_\textrm{2D}]$. Because the electron and hole distribution do not have dipole components,  we set $\rho_\textrm{1M}=1$, $\rho_\textrm{1D}=0$ and $\phi_\textrm{2M}=1$, $\phi_\textrm{2D}=0$.

It is useful to define an effective dielectric function for the electron-hole interaction as the unscreened Coulomb interaction over the screened one:
\begin{equation}
\label{eq:eps_q_vdWH}
\epsilon(q_\parallel) = \frac{\underline{\rho}_{\hspace{0.05cm}\textrm{e}}^{\intercal}(q_\parallel)~\underline{\phi}_{\hspace{0.05cm}\textrm{h}}(q_\parallel)}{\underline{\rho}_{\hspace{0.05cm}\textrm{e}}^{\intercal}(q_\parallel)~\underline{\underline{\epsilon}}^{-1}(q_\parallel)~\underline{\phi}_{\hspace{0.05cm}\textrm{h}}(q_\parallel)},
\end{equation}

\begin{figure*}[!t]
 \includegraphics[width = 0.7\linewidth]{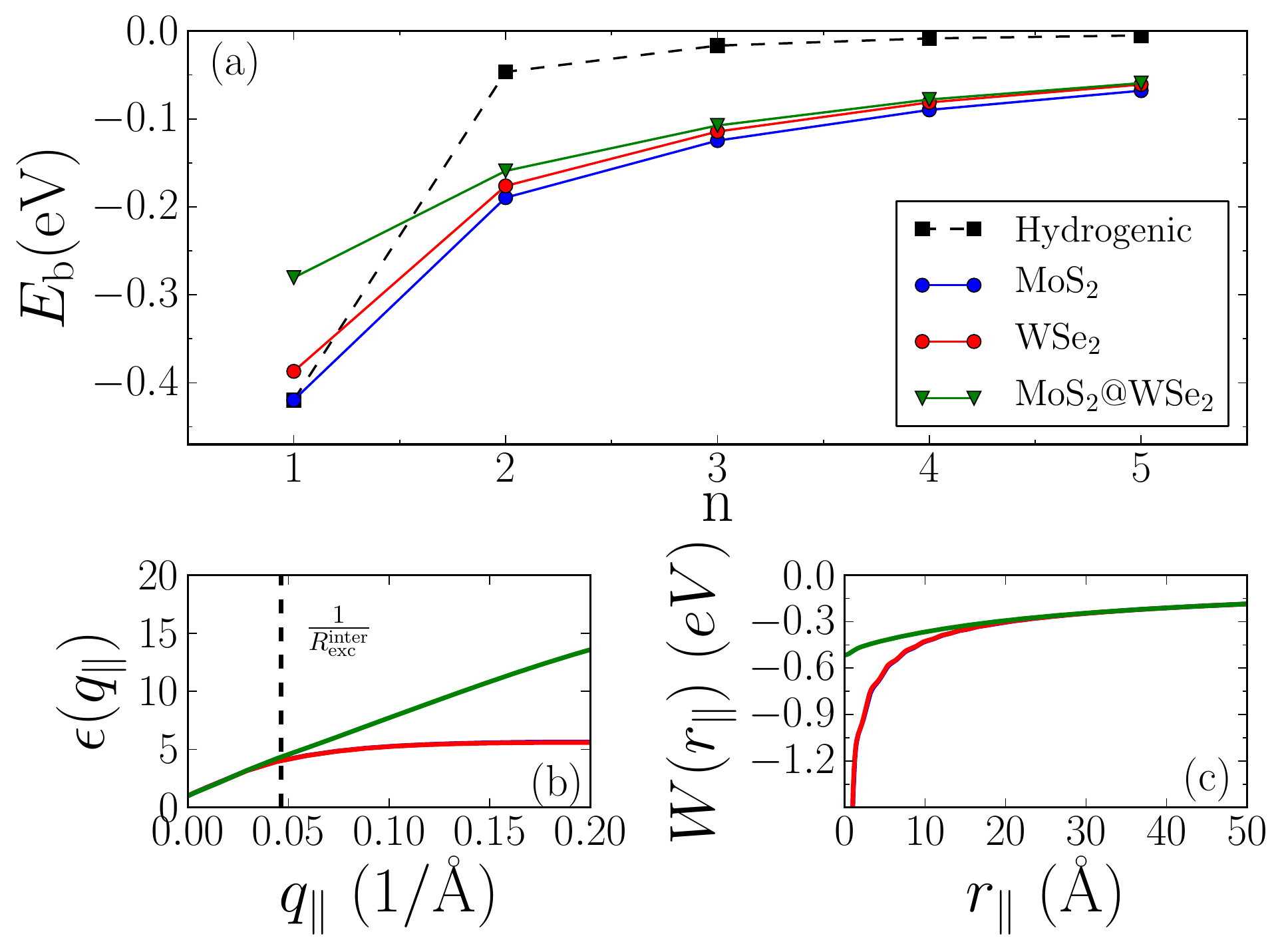}
  \caption{Panel (a) Rydberg series for intra and interlayer excitons. The hydrogenic series is obtained from the standard expression for the energy of a 2D hydrogen atom, i.e. $E_\textrm{b}= \frac{\mu_{ex}}{2(n-1/2)^2\epsilon^2}$ and by fitting $\epsilon$ to the lowest lying excitonic state in MoS$_2$.  Panel (b) effective dielectric function and panel (c) Screened electron-hole interaction for intra and interlayer excitons in bilayer MoS$_2$-WSe$_2$. }
  \label{fig:RydbergSeries}
\end{figure*}

A typical signature of excitons in two-dimensional materials is the non-hydrogenic Rydberg series\cite{Chernikov2014}. The Rydberg series along with the full wave-vector dependent effective dielectric screening and electron-hole interaction for the intra and inter-layer excitons in the MoS$_2/$WSe$_2$ bilayer are shown in \cref{fig:RydbergSeries} (a), (b) and (c) respectively.
The non-hydrogenic nature of the Rydberg series should be clear from the comparison with the dashed line in panel (a) which represents the Rydberg series obtained from an hydrogenic equation where the electron-hole interaction is screened by a constant dielectric function. The first interesting characteristic to notice is that the intra and interlayer Rydberg series converge towards each others for higher excited excitonic states. This is not surprising considering that higher lying states are more delocalized and once their radius is much greater than the heterostructure thickness intra and interlayer excitons are practically indistinguishible, provided that the screening of the electron-hole interaction  is comparable.  As shown in panel (b), the effective dielectric function is indeed the same for inter and intra-layer excitons within the region of relevant wavevectors values, i.e. values of $q_\parallel$ smaller than the reciprocal of the exciton radius (indicated with a vertical line in panel (b) for the lowest lying exciton). However, it is clear from panel (c) that the screened interaction for the interlayer exciton is lower than for the intralayer ones, and it does not diverge for $r_\parallel\rightarrow 0$. This is a simple consequence of the finite electron-hole spatial separation, which guarantees that the electron and the hole are separated even for $r_\parallel = 0$.

\begin{figure*}[!b]
 \includegraphics[width = 0.65\linewidth]{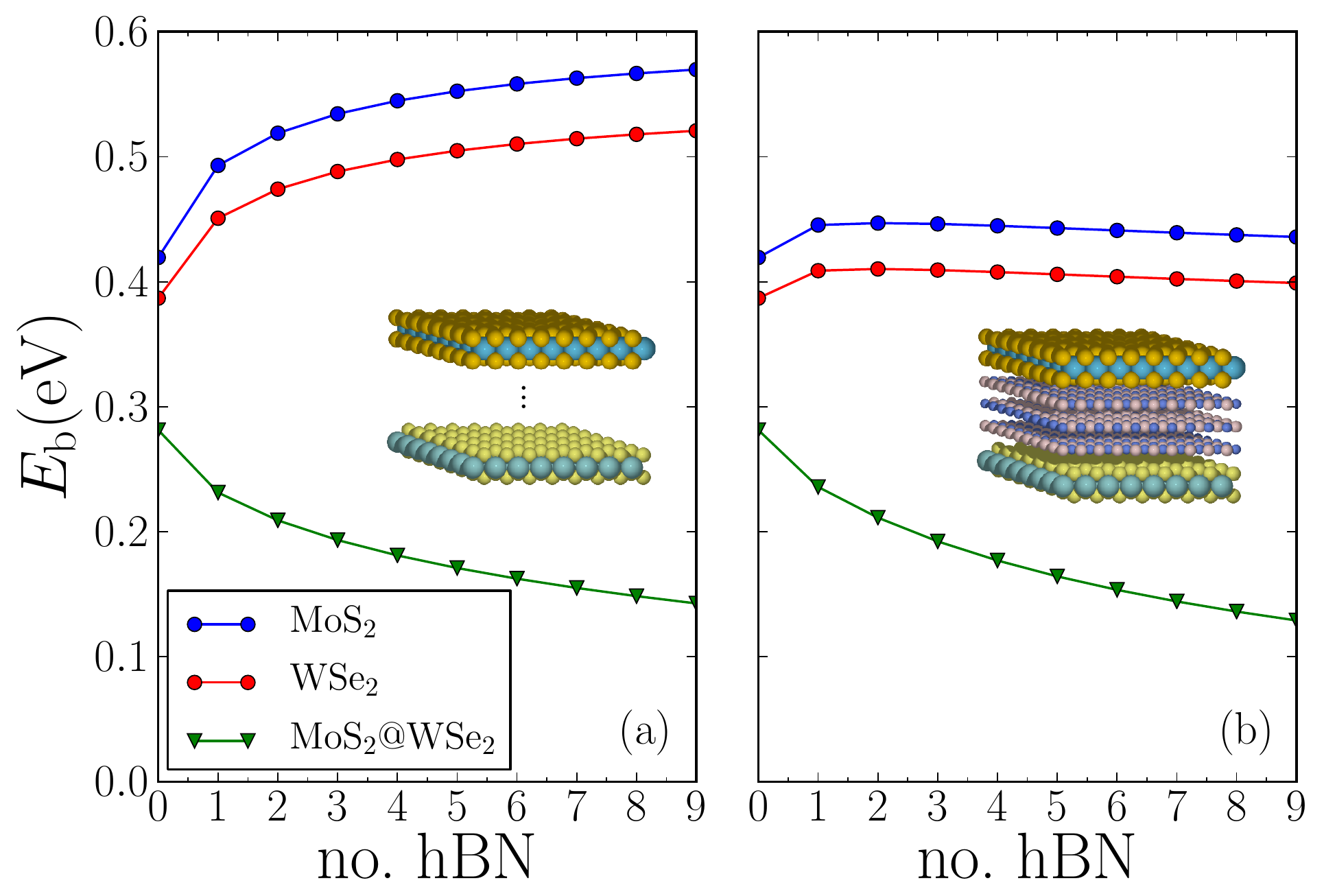}
  \caption{Intra and Interlayer exciton binding energy as a function of number of intercalating vacuum (a) and h-BN (b) layers.}
  \label{fig:distance_dep}
\end{figure*}

To explore the effect of spatial separation even further, we study the case of MoS$_2$-WSe$_2$ heterostructures intercalated with h-BN. To isolate the effect of screening, we perform the same calculations with h-BN is substituted by vacuum. The results for the lowest intra and inter-layer exciton binding energies are shown in panel (a) and (b) of \cref{fig:distance_dep}.
Clearly the behavior of the interlayer exciton is quite similar in the two cases, meaning that the main effect of inserting hBN layers is to increase the electron-hole separation. In contrast, the binding energy of the intralayer excitons increases in the case of increasing vacuum while it remains constant when more hBN layers are inserted. This is because moving MoS$_2$ and WSe$_2$ apart decreases the screening whereas inserting more hBN layers increases it, leading to an overall compensation. \\

When stacking 2D layers together, the exciton binding energy is not the only quantity that is affected. Indeed, also the band gap of each of the layers in the stack is reduced due to the increased dielectric screening. The state of the art method for properly including the effect of dielectric screening in band structures is the G$_0$W$_0$ method\cite{Cheiwchanchamnangij2012,Huser2013,Falco2013,Filip2015}. In this many-body theory based approach, the information of electronic screening is contained in the dynamical screened Coulomb potential:
\begin{equation}
\bar{W}_{{\bf GG}'}({\bf q},\omega ) = \left[\epsilon^{-1}_{{\bf GG}'}({\bf q},\omega ) - \delta_{{\bf GG}'}\right]\frac{4\pi}{|{\bf G}+{\bf q}|^2}.
\end{equation}
Computing $\bar{W}_{{\bf GG}'}({\bf q},\omega )$ is a demanding task even for simple materials and it is practically impossible for multi-layer vdWHs. Fortunately, as described elsewhere \cite{Kirsten2017}, the effect of screening on the band structure of a given layer in a van der Waals stack, can be accounted for by combining the QEH model with a standard G$_0$W$_0$ method at the computational cost of a monolayer calculation. The main idea is to correct $\bar{W}_{{\bf GG}'}({\bf q},\omega ) $ for a given layer in the following manner:
\begin{equation}
\bar{W}^\textrm{vdWH}_{{\bf GG}'}({\bf q},\omega ) = \bar{W}^\textrm{monolayer}_{{\bf GG}'}({\bf q},\omega ) +\Delta W({\bf q},\omega )\delta_{{\bf G}0}\delta_{{\bf G}'0}
\end{equation}
where $\Delta W({\bf q},\omega )$ is the correction to the head of the matrix $\bar{W}_{{\bf GG}'}$ that includes the extra screening coming from the neighboring layers. For a given layer, such a correction is efficiently calculated within the QEH model. Indeed, by using an expression equivalent to \cref{eq:W_q_vdWH}, the electron-electron interaction is calculated for the isolated layer and the layer in the vdWH, then $\Delta W({\bf q},\omega )$ is obtained as the difference between the two interactions. Once corrected, the screened potential can be used directly in a standard monolayer G$_0$W$_0$ calculation.  
With this approach, which we refer to as G$_0\Delta$W$_0$, we are able to efficiently calculate the band positions of any vdWHs and, specifically for this work, the position of the valence band maximum and conduction band minimum of MoS$_2$ and WSe2$_2$ for a varying number of hBN layers. A verification of the G$_0\Delta$W$_0$ approach for the specific MoS$_2/$WSe$_2$ system is provided in the supporting information.

The level alignment for a bilayer MoS$_2$-WSe$_2$ obtained from the G$_0\Delta$W$_0$  and including spin-orbit coupling effects is shown in \cref{fig:model_sketch} (a) in black lines. To highlight the effect of interlayer screening we reported, in the same panel, the band edges for the isolated monolayers calculated within the G$_0$W$_0$ approximation (colored bars). \Cref{fig:model_sketch} (b), instead, illustrates the difference in intra and interlayer gaps with respect to the isolated layers as a function of the number of intercalated hBN layers.
Although the band gap renormalization is noticeable when going from monolayers to bilayer, the intercalation of h-BN does not have a considerable effect. Comparing to the variation in exciton binding energy in panel (c), we observe a similar trend for the intralayer cases whereas for the interlayer case the trend is different. This is reasonable since, differently from the interlayer exciton, there is not explicit dependence of the interlayer gap on MoS$_2$-WSe$_2$ separation. 

%
With the knowledge of band edges position and exciton binding energies we are now ready to calculate the position of the excitonic photoluminescence (PL) peaks in MoS$_2$/WSe$_2$ based heterostructures.

\begin{figure*}[!t]
  \includegraphics[width = 0.9\linewidth]{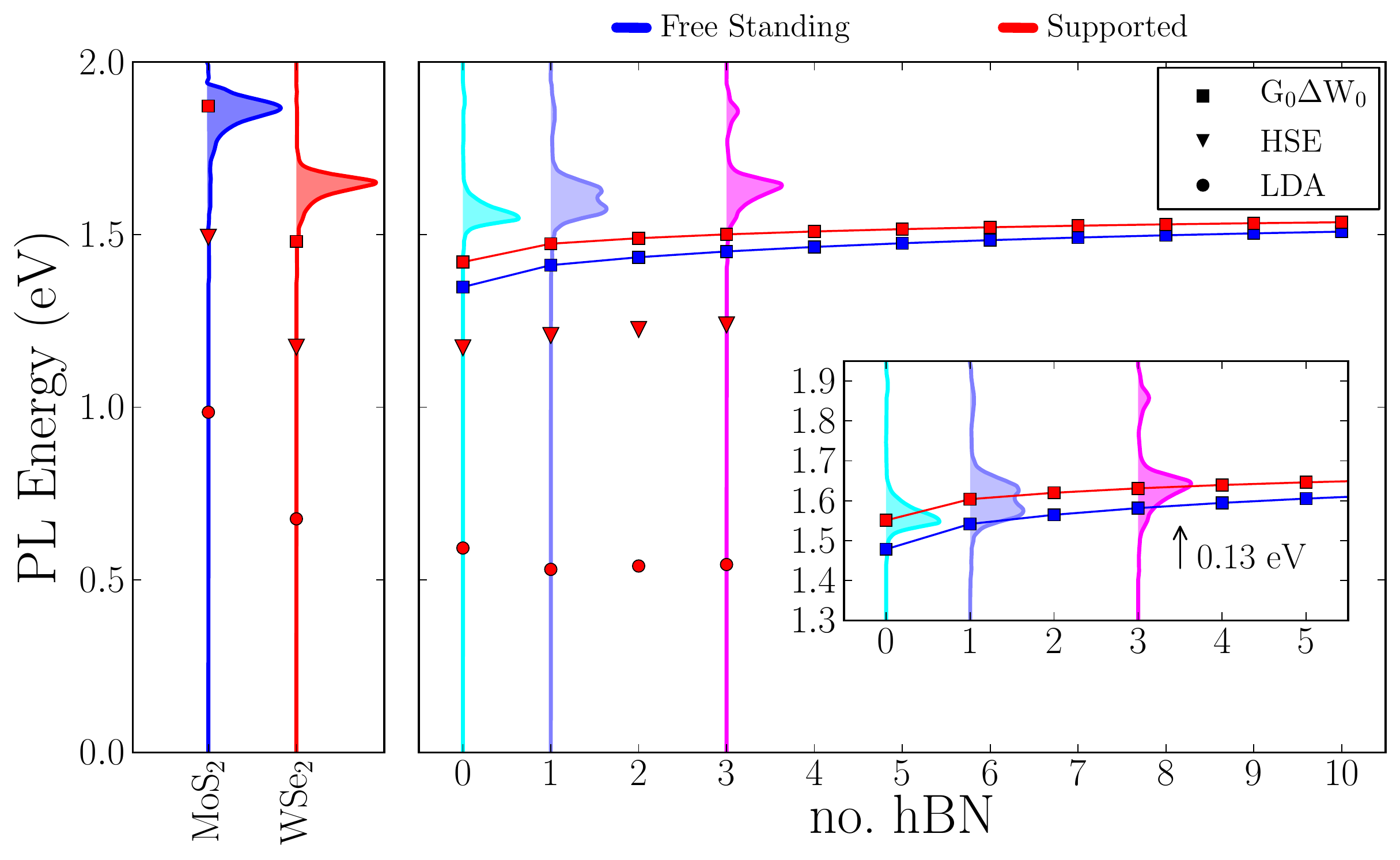}
  \caption{Comparison of the calculate position of the excitonic photoluminescence peaks with experimental data for isolated layers (left panel) and MoS$_2$-WSe$_2$ based heterostructures (right panel). The experimental data\cite{Fang2014} is reported as shaded colored curves. Inset: same as in the main figure, but with a shift of $0.13$ eV, to highlight that we can well reproduce the trend.}
  \label{fig:indirect_peak}
\end{figure*}

The photoluminescence signal is generated by radiative electron-hole pairs recombination. Considering that typical radiative recombination times in TMDCs are much longer than electron and hole thermalization times, we expect the exciton recombination to happen from the K-point of the conduction band in MoS$_2$ and the K-point of the valence band in WSe$_2$. We note that, the lattice mismatch and a non-zero alignment angle between the two layers implies a mismatch of the first BZ of the two materials, as shown in \cref{fig:bands_rotation}. This means that for a radiative transition to happen the momentum mismatch has to be compensated by some other physical mechanism.  Mechanisms of this kind could include phonon assisted transition, electron-electron interaction, defect scattering or breaking of momentum conservation induced by the exterior potential field generated by the neighboring layers\cite{Yu2015}. Here we focus on the energetics of the process, which should not be effected by the particular recombination mechanism. 
Taking into account the type-II band alignment, the position of the pholuminescence peak of the lowest bound exciton is given by:
\begin{equation}
E_\textrm{PL} = E_\textrm{IG}-E_\textrm{b}^\textrm{Inter}
\end{equation}
where $E_\textrm{IG}$ is the interlayer electronic gap and $E_\textrm{b}^\textrm{Inter}$ the interlayer exciton binding energy.
The positions of the lowest energy photoluminescence peak for isolated layers, as well as for MoS$_2$-WSe$_2$ based heterostructures with a varying number of intercalating hBN layers are plotted in \cref{fig:indirect_peak} for the free standing and supported case. Experimental photoluminescence spectra from Ref.~\citenum{Fang2014} are also reported in the same figure. For the supported case we use $30$ layers of hBN to simulate the effect of a substrate. The choice of hBN as a substrate is enforced by the QEH approach which applies only to layered materials, but it is justified by the fact that hBN has a bulk dielectric constant similar to SiO$_2$, which is the substrate used in the experiment.
The agreement with the photoluminescence peaks for MoS$_2$ it is good but for WSe$_2$ it is underestimated by around $0.13$ eV. Roughly the same constant shift is seen for the heterostructures. This indicates that the deviation is due to a too high positioning of the WSe$_2$ valence band by the G$_0\Delta$W$_0$. However, the agreement is still highly satisfactory. Indeed shifting our values by $0.13eV$ we can well reproduce the data for increasing number of intercalating layers as shown in the inset. This indicates that it is the estimated indirect bandgap to be slightly off, but a difference of just $0.13$ eV is still quite accurate for our multi-step approach. According to the experimental interpretation in Ref.~\citenum{Fang2014}, while the PL peak for the bilayer without hBN shows a clear interlayer exciton peak, with the intercalation of hBN the interlayer exciton signal is reduced and eventually covered by the WSe$_2$ intralayer exciton PL peak for three h-BN layers. If this is the case, it is not possible to completely validate the trend of our ab-initio PL peaks values through the actual experimental data. Anyways to demonstrate that using G$_0$W$_0$ for monolayer bands is strictly necessary, \cref{fig:indirect_peak} shows the photoluminescence peaks obtained using a band alignment from LDA and HSE calculations. It is evident that the LDA dramatically underestimates the position of the photoluminescence peaks for both the isolated layer and the bilayer case. The HSE improves the band alignment significantly, but is still around $0.3$ eV below the experimental values. Furthermore even the trend of the indirect exciton peak as a function of hBN layers is reversed by LDA. This is a consequence of the strong charge transfer predicted by LDA as discussed earlier. Indeed, charge transfer tends to open the indirect band gap and therefore shift the PL peaks up in energy, with a shift that decreases with increasing number of hBN layers. This shift in energy is larger and opposite to the optical band gap reduction due to interlayer excitons, which explains why the position of the LDA PL peaks descrease in energy in contrast with the G$_0$W$_0$ results.

In conclusion, we presented a general approach to calculate band alignment and interlayer excitons in incommensurate van der Waals heterostructures. For the MoS$_2$/WSe$_2$ heterostructure, we found that interlayer hybridization is important only around the $\Gamma$-point of the BZ and therefore does not influence the opto-electronic properties that are governed by states around the K-point. This implies that an accurate description of the band edge positions can be obtained from the isolated monolayer band structures aligned relative to a common vacuum level and renormalized by the polarization effect from neighboring layers.  In particular, we determined the effect of interlayer polarization to account for $\sim 8 \%$ reduction in the intelayer gap. On the other hand, the effect on band alignment of the formation of an interface dipole is expected to be negligible for the studied system. We find interlayer excitons to have significant binding energies of up to $0.3$ eV and showing monotonic decrease with the layer separation. Comparison with experimental photoluminescence spectra revealed a constant redshift of the calculated lowest optical transition of around $0.13$  eV, which we ascribe to a slight overestimation of the WSe$_2$ valence band edge by G$_0$W$_0$. This indicates that for describing photoluminescence it is crucial to obtain accurate band edges at the isolated layers level. 
Finally, our calculations show that it is possible to obtain quantitatively accurate band- and exciton energies for rather complex vdWHs when employing proper methods, and highlight the deficiencies of standard density functional theory for band alignment problems.  

\section{Methods}
\label{sec:meth}
All the ab-initio calculation in this work are performed with GPAW\cite{Enkovaara2010,GPAW_web}.  The band structures of the twisted bilayers were calculated at the DFT level with an LDA exchange correlation functional and double-zeta polarized atomic orbitals as a basis set. The HSE06 and LDA calculations for the monolayers were performed using a plane wave basis set with a cut off energy of $500$ eV and $18\times18$ k-point grids.

We find it useful to elaborate on the color scheme choice for the twisted bilayer band structure in \cref{fig:bands_rotation}. Such choice should be understood by considering that the two layers have different primitive cells and that the unfolding procedure has to be performed separately. Indeed, the unfolding for a given layer, say MoS$_2$, not only will project the eigenstates belonging onto MoS$_2$ to their primitive cell but also the WSe$_2$ ones. As the latter should rather be projected onto the WSe$_2$ primitive cell, we ``hide'' them by choosing a color scheme that goes from blue to white and decreases the size of the markers when going from states localized completely in  MoS$_2$ to states localized completely in WSe$_2$. The same argument applies to the unfolding to the WSe$_2$ cell, but  a red color scheme is applied instead.

For lattice matched heterostructures modeled in a minimal unit cell, it was checked that the atomic orbital basis yields the same band structure as well converged plane wave calculations.  
For the calculation of dielectric properties of van der Waals heterostructures we utilized dielectric building blocks available in Ref.~\citenum{database2015}. Specifically the response function of each building block was calculated on a plane-waves basis with $100$ eV cut-off energy and $100\times100$ k-point mesh. In order to avoid spurious interaction from artificial replica in the out-of-plane direction, a truncated Coulomb interaction with $20$ \AA~ of vacuum is used. The interlayer distance between the layers are taken as average of the interlayer distance in their respective bulk form, specifically $d_{\textrm{MoS}_2/\textrm{WSe}_2}=6.51$\AA, $d_{\textrm{MoS}_2/\textrm{hBN}}=5.08$ \AA~, $d_{\textrm{hBN}/\textrm{hBN}}=3.2$ \AA~ and $d_{\textrm{WSe}_2/\textrm{hBN}}=5.28$ \AA. The monolayer G$_0$W$_0$ calculations have been performed employing a new efficient technique\cite{Rasmussen2015} that overcomes the problem of slow convergence of the band structures with k-point grid and yields well converged band gaps with $18\times18$ k-points (rather than $40\times40$ using standard approaches). We used an energy cut-off of $150$ eV for the dielectric function and sum over empty states. The G$_0$W$_0$ band energies were extrapolated as $1/N_G$ to the infinite plane wave limit.

The Mott-Wannier equation in \cref{eq:MW} was solved on a radial logarithmic grid ensuring numerical convergence of exciton energies up to $0.002$ eV.

\section{Acknowledgement}
The authors acknowledge support the Center for Nanostructured Graphene (CNG), which  is sponsored by the Danish National Research Foundation, Project DNRF58. The authors declare no competing financial interests.

\bibliography{}
\pagebreak

\section{Supporting Information}

\subsection{Effect of Hybridization and Charge-Transfer}
In the main text we argued that the use of supercells is essential for a good description of the band structure of mismatched bilayers and the main differences between the bands of the isolated layers and the bilayers are consequence of charge transfer and hybrization separately. In this section of the supporting information we prove these arguments for the case of MoS$_2/$WSe$_2$ based structures.

\begin{figure*}[!h]
  \includegraphics[width = 0.8\linewidth]{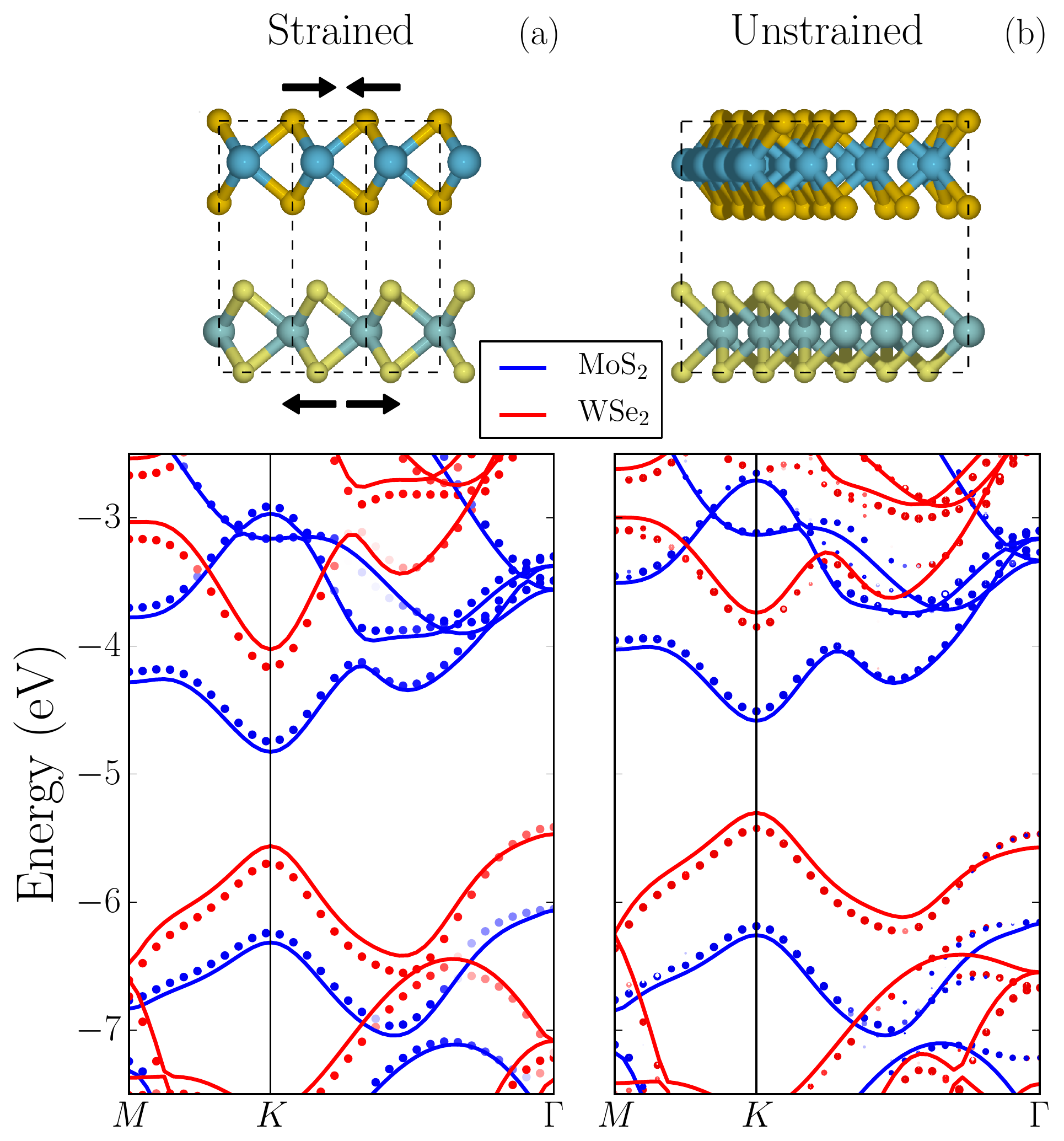}
   \caption{Panel (a) electronic bands for strained bilayer  MoS$_2/$WSe$_2$ (circles) and isolated monolayers (continuous line). Panel (b) the same as in (a) but for the unstrained structures. For the unstrained bilayer an alignment angle of $\sim16.1^{\circ}$ is used. In the strained structures the lattice parameter is the average of the lattice parameter of isolated MoS$_2$ and WSe$_2$. The figure shows that the effect of charge transfer can be inferred from a simpler strained calculation.}
 \label{fig:ct_strain}
\end{figure*}

We start out by demonstrating that using a supercell is unavoidable if an accurate band structure of the MoS$_2/$WSe$_2$ is needed. This is shown in fig.~\ref{fig:ct_strain}, where we plot the band structure of the strained bilayer and isolated monolayers in panel (a) and the corresponding unstrained structures in panel (b).
In both panels, the bands belonging to the isolated layers are drawn with continuous lines while the ones for the bilayers are drawn with circles. The unstrained bilayer is constructed using a supercell and an alignment angle of $\sim16.1$ as described in the main text, whereas for the strained bilayer we use a unit cell with the lattice parameter equal to the average of the lattice parameter of the isolated monolayers. From the figure it is evident that straining the layers has a considerable effect both on the curvature of the bands and on their positioning with respect to vacuum. We thus conclude that accurate band structures cannot be obtained without employing supercells. 
However, it is still possible to extract information about charge transfer and hybridization from the strained calculations. Indeed, we can see from fig.~\ref{fig:ct_strain} that the relative difference between isolated layers and bilayer, in both panels, are practically the same. 

\begin{figure*}[!b]
  \includegraphics[width = 0.8\linewidth]{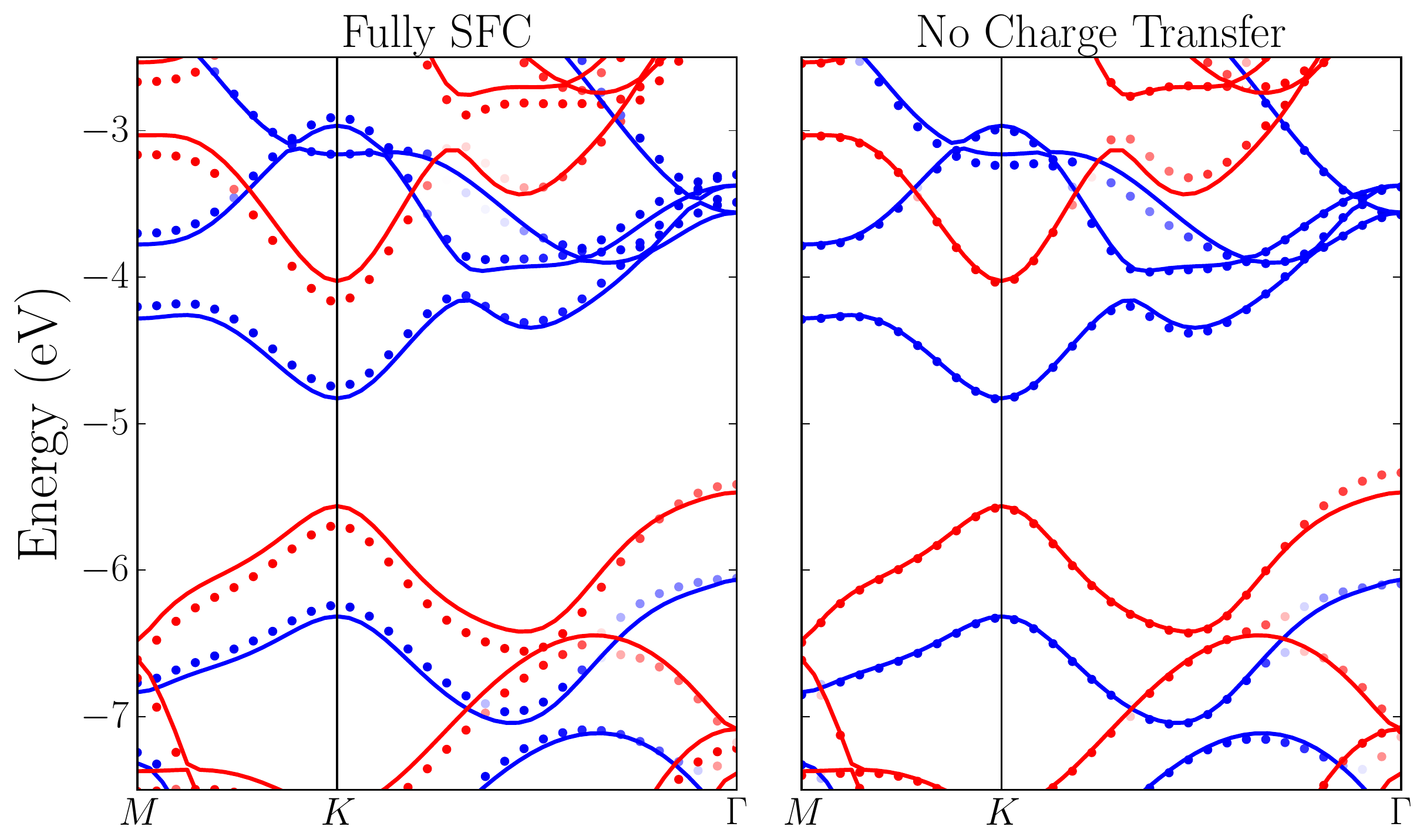}
   \caption{Left panel: bands for strained MoS$_2/$WSe$_2$ bilayer (circles) and constituent monolayers (continuous lines). Here the charge transfer effect manifests as constant shift in opposite direction for MoS$_2$ (in blue) and WSe$_2$ (in red). Right panel bands for the same systems but for the bilayer the Hartree potential is not updated self-consistently. Keeping the Hartree potential fixed to the one of the isolated layers prevent charge transfer and therefore bilayer and monolayers bands coincide as long as hybrization is negligible. Hybridization is present around $\Gamma$.}
 \label{fig:noct}
\end{figure*}
Based on this consideration we proceed the analysis of charge transfer and hybridization using the strained calculations, which are computationally more feasible. As explained in the main text, we observe a constant shift in energy, upwards for MoS$_2$ and downwards for WSe$_2$, and a wavevector dependent variation around $\Gamma$ when comparing the isolated layers to the bilayer. While the effect of hybridization is a direct consequence of the mixing among wavefunctions of the two layers, charge transfer results from the rearrangement of the electrons at the bilayer interface due to the difference in band gap centers of the two materials. From a DFT calculation point of view, it is the self-consistent procedure, in particular the change in the Hartree potential in each loop, that allows the rearrangement of the electrons once the two materials are put together. This means that performing a non-self-consistent DFT calculation starting from the self-consistent ground state density of the isolated layers, would not allow for the update of the Hartree potential and consequent electrons rearrangement. The fully self-consistent band structures (reported for comparison from fig.~\ref{fig:ct_strain} (a)) and the non self-consistent ones are shown in fig.~\ref{fig:noct}, left and right panel respectively.
In panel (b), the isolated layers bands are now exactly on top of the bilayer ones throughout most of the Brillouin zone and therefore it should be now clear that the rigid shift of the bands was a signature of charge transfer. Furthermore the alteration of the bands around the $\Gamma$ point has not disappeared. This is exactly what we expected considering that hybrization is a result of the overlap of the wavefunctions which is accounted for in the non-self-consistent calculation.

\subsection{Validity of the QEH correction on G$_0$W$_0$ band structure}
\begin{figure*}[!h]
  \includegraphics[width = \linewidth]{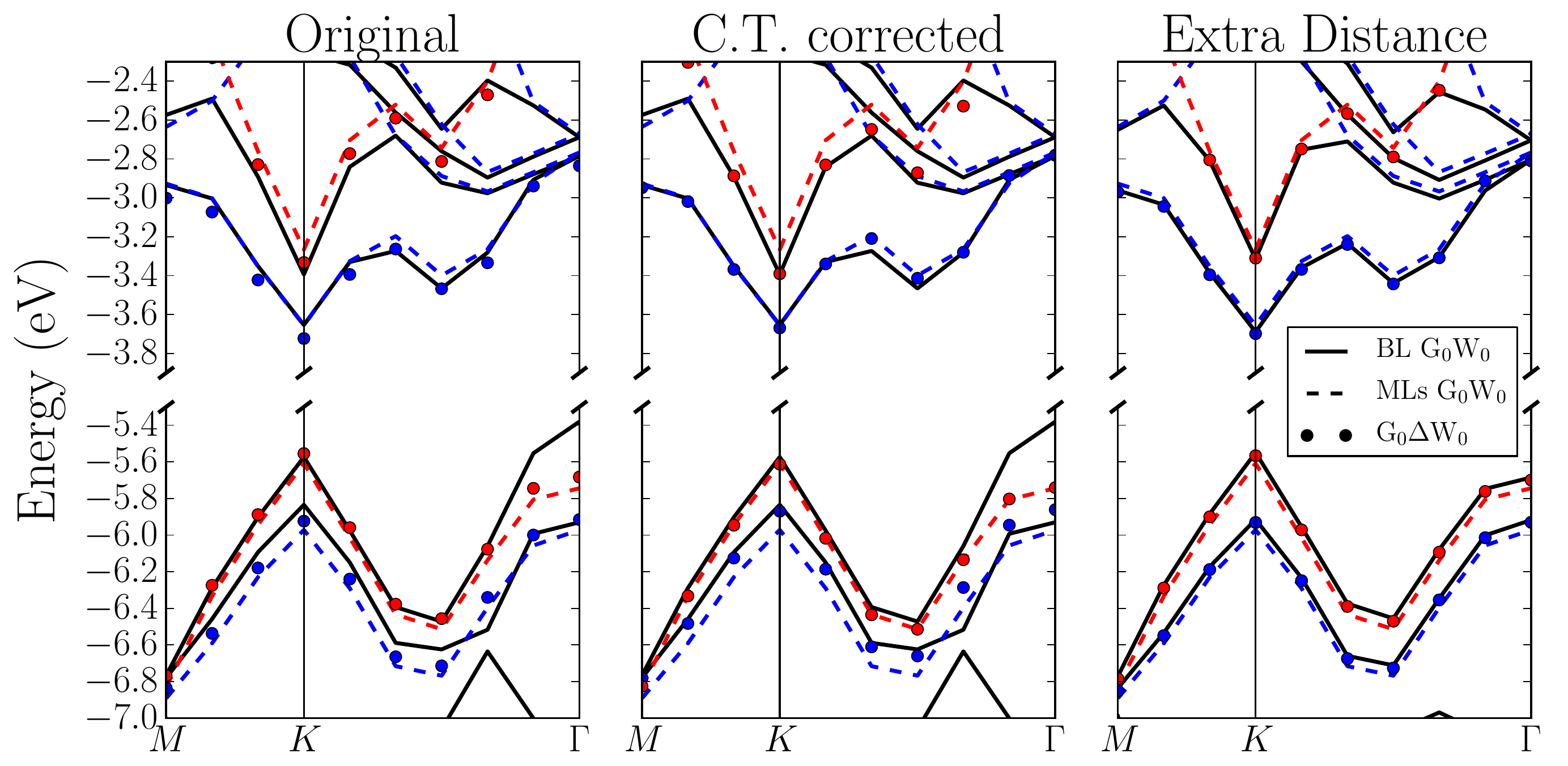}
   \caption{Comparison of the G$_0\Delta$W$_0$ method to G$_0$W$_0$ calculation for the lattice-matched bilayer. Left panel: reference calculation. Central panel:  G$_0\Delta$W$_0$ bands are shifted upwards for MoS$_2$ and downwards for WSe$_2$ to account for the charge transfer. The values of the shifts are extracted from the comparison between the isolated layers and bilayer LDA calculations. Right panel: the layer separation is artificially increased by $3$\AA. With the extra distance we expect the effect of charge transfer and hybridization to be completely negligible.The isolated layer bands are reported in all the panels as reference. As usual blue is used for MoS$_2$ and red for WSe$_2$.}
 \label{fig:test_QEH}
\end{figure*}

To check the validity of our G$_0\Delta$W$_0$ approach we perform G$_0$W$_0$ calculations for strained MoS$_2$ and WSe$_2$ isolated layers and MoS$_2/$WSe$_2$  bilayers. The choice of strained structure is obviously imposed by the feasibility of a G$_0$W$_0$ calculation for the bilayers. For the following calculation plane-wave mode has been used. In the left panel of fig.~\ref{fig:test_QEH} we report the bands for the strained  MoS$_2/$WSe$_2$ bilayer from a G$_0$W$_0$ calculation (continuous black line) and the G$_0$W$_0$ approach.
The G$_0$W$_0$ bands for the isolated layers are also shown as reference. As expected from the extra screening that each layer provides to the other, the intra and inter layer gaps are reduced compared to the isolated layer ones and such an effect is grasped both from the full G$_0$W$_0$ calculation and the G$_0\Delta$W$_0$ method. However, the agreement between G$_0$W$_0$ and G$_0\Delta$W$_0$ is not striking. This is because the effect of charge transfer is still present at the G$_0$W$_0$ level, since the Hartree potential generated by the charge rearrangement at the interface is the same as the DFT one. Only self-consistency, indeed, could relieve this problem. To prove that charge transfer is still there and that it is an effect inherited from the starting LDA calculation, we evaluate the layer dependent energy shift at the $K$-point by comparing isolated layers and bilayers bands at the LDA level and then add these shifts to the G$_0\Delta$W$_0$ bands. The results are shown in the central panel of fig.~\ref{fig:test_QEH}. The agreement is nearly perfect and it supports our argument on the importance of charge transfer. As a side note, we mention that the effect of charge transfer using plane-wave mode, as opposed to LCAO, is a bit lower, namely we get an increase in interlayer gap of $0.11$ eV compared to the $0.21$ eV reported in the main text.

As a further proof of the validity of the G$_0\Delta$W$_0$ method we repeat the G$_0$W$_0$ for the bilayer adding $3$ \AA~ to interlayer distance between MoS$_2$ and WSe$_2$. This guarantees that charge transfer and hybridization effects are negligible. Screening effects, on the other hand, are still appreciable being the Coulomb coupling between the layers long range. The bands for such a system are shown in the right panel of fig.~\ref{fig:test_QEH} and it is clear that the G$_0\Delta$W$_0$ does a good job.

\end{document}